\documentclass{llncs}

\usepackage{oz}
\usepackage{times}
\usepackage{url}
\usepackage{alltt}
\usepackage{graphicx}

\newlength{\zedwidth}
\def\plus{{^+}}
\def\star{{^*}}

\begin {document}


\def\fuzz{{\large\it f}{\sc uzz}}

\def\emquote{\vspace{-.8ex}\begin{quote}\small\em}
\def\endemquote{\end{quote}\vspace{-.8ex}}

\def\zsidebyside{\begin{sidebyside}}
\def\endzsidebyside{\vspace{-2ex}\end{sidebyside}}

\mainmatter

\title{Online Communities: Visualization and Formalization}

\author{Jonathan P.\ Bowen}
\institute{
Museophile Limited, Oxford, UK \\
\email{jonathan.bowen@lsbu.ac.uk} \\
\url{www.jpbowen.com}
}

\date{}

\maketitle

\begin{abstract}
Online communities have increased in size and importance
dramatically over the last decade. The fact that many
communities are online means that it is possible to extract
information about these communities and the connections between
their members much more easily using software tools, despite
their potentially very large size. The links between members of
the community can be presented visually and often this can make
patterns in the structure of sub-communities immediately
obvious. The links and structures of layered communities can
also be formalized to gain a better understanding of their
modelling.  This paper explores these links with some specific
examples, including visualization of these relationships and a
formalized model of communities using the Z notation.  It also
considers the development of such communities within the {\em
Community of Practice} social science framework.  Such
approaches may be applicable for communities associated with
cybersecurity and could be combined for a better understanding
of their development.
\end{abstract}

%

\section{Introduction}
\label{introduction}

The development of collective human knowledge has always
depended on communities.  As communities have become more
computer-based, it has become easier to monitor the activity of
such interactions \cite{Don95}.  Recently the increasing use of
online communities by the wider population (e.g., for social
networking) has augmented the ways that communities can form and
interact since geographical co-location is now much less
critical than before the development of the Internet and the web
\cite{Bel04,Bor11}.

Here, we consider the visualization of online
communities, their development in a {\em Community of Practice}
(CoP) context \cite{Bow11,Wen02}, their formalization using the Z
notation \cite{Spi01}, and possible applications to
cybersecurity.

\section{Visualization}
\label{visualization}

A community of people can be modelled naturally as a
mathematical graph with vertices representing people and edges
representing connections between those people. This is not
dissimilar to the web with its pages and hyperlinks \cite{Num05}.
The edges in the graph may be undirected (e.g., for friendship
between two people where both like each other, or as
collaborators in some joint activity such as co-authorship of
joint publications \cite{Bow13,Bow12}) or directed (e.g., for a
citation of one author to another author's work). Such a graph
is a natural way to visualize relationships between people. When
observed visually, patterns in the graph can be quickly
assimilated and analyzed by the viewer.

\begin{figure}[htp]
\begin{center}
\includegraphics[width=\textwidth]{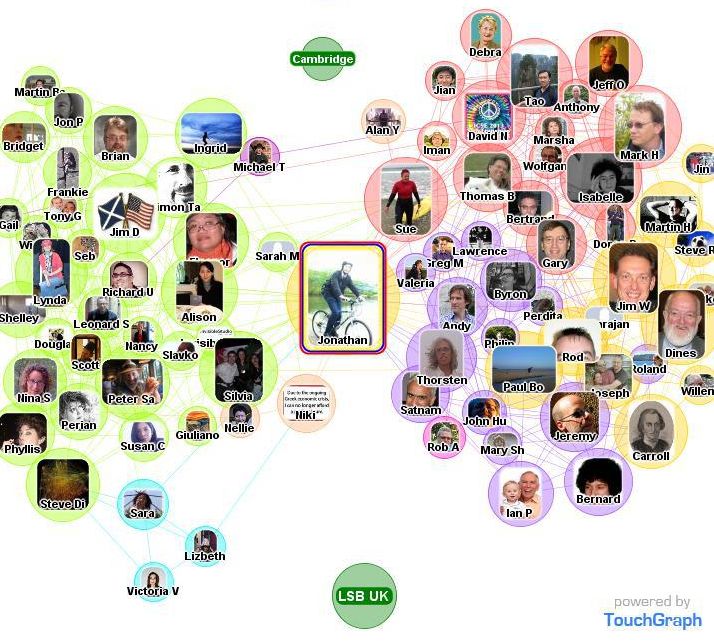}
\end{center}
\caption{
Facebook TouchGraph connections.
\label{facebook}}
\end{figure}

With the increase in social and professional networking online,
the visualization of online communities in an automated way as
graphs is now relatively easy. For example, Figure
\ref{facebook} shows connections between one of the authors and
``friends'' on Facebook, using visualization software provided
by TouchGraph (\url{http://www.touchgraph.com}). The TouchGraph
Facebook app tool also takes account of links between all the
people included in the graph
(\url{https://apps.facebook.com/touchgraph/}). Thus it is
possible to note groups within the network visually.  Greatly
interconnected groups of people are clustered together on the
displayed graph and are highlighted using colours. For example,
in the case of Figure \ref{facebook}, those towards the right of
the diagram are mainly people interested in computer science and
those towards the left are mainly interested in museums and the
arts, two major but largely non-overlapping areas of interest to
the author (in the centre of this diagram). Within the computer
science community, several sub-communities are indicated by
different colours.

\begin{figure}[htp]
\begin{center}
\includegraphics[width=\textwidth]{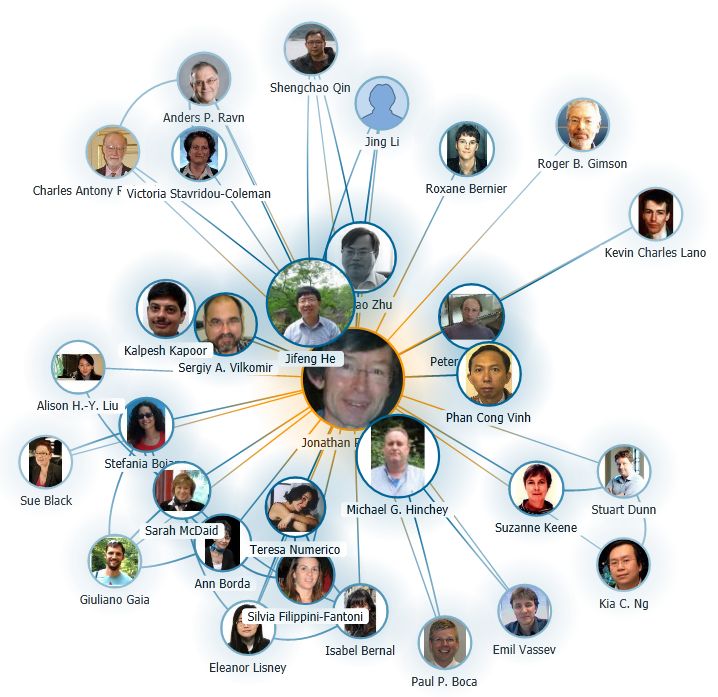}
\end{center}
\caption{
Academic Search co-authorship connections.
\label{coauthors}}
\end{figure}

Figure \ref{coauthors} shows a similar set of links for
co-authors, provided by Microsoft Academic Search
(\url{http://academic.research.microsoft.com}) as part of its
visualization toolset. Here the linking of co-authors in the
fields of formal methods and museum informatics can also be
observed as linked clusters mainly to the top and bottom of the
displayed graph respectively.

Similar visualization tools could be applied to a wide variety
of communities, especially if the activities of an indivisual
are under investigation (e.g., for cybersecurity reasons). Foe
example, patterns of email connectivity could be displayed in a
similar manner.

\section{Communities of Practice}
\label{CoP}

Increasingly it has been realised that the development of
communal knowledge is largely social in character, although it
often takes place in a variety of situations, whether in the
workplace or some other organization.  Such social
considerations have also led to the theoretical framework of a
{\em Community of Practice} (CoP) in the social sciences field,
with a number of elements, principles, and developmental stages
\cite{Hug07,Wen98,Wen02}.  A CoP is a group of people with a shared
interest or profession, engaged in the developing communal
knowledge. It involves situated learning, in which the people
that are learning also apply this knowledge in the same context
(e.g., during practical experience).

The following three fundamental elements form the structural
model of a community of practice \cite[chapter 2]{Wen02}:

\begin{enumerate}

\item
{\em Domain:} A CoP must have a common interest to be effective. All
the participants in the group must be able to contribute in some
way within this domain. Otherwise it is just a collection of
people with no particular purpose. For example, cybersecurity is
such a domain. Formal methods and visualization techniques are
other examples.

\item
{\em Community:} A CoP also needs a group of people who are
willing to engage with at least some others in the group, so
ultimately the entire group is transitively connected as a
single entity, from a global viewpoint. This aspect is critical
to the effective development of knowledge. The group of people
interested in cybersecurity includes both those attempting to
protect and break into systems, with some people crossing the
the boundary between these two aspects (often as ``poachers
turned gamekeepers''). This community has expanded rapidly with
the rise of the Internet and the web.

\item
{\em Practice:} The CoP must explore both existing knowledge and
develop new knowledge, based on existing concepts, but expanded
through actual application in a practical sense. This leads to a
set of common approaches and shared standards in applying them.
In the case of cybersecurity, this includes two specific
sub-communities, those wishing to increase the effectiveness of
cybersecurity and those attempting to break it.

\end{enumerate}

Developing a successful CoP requires the interplay of these
three elements within a community in a balanced manner, because
they are all dynamically changing over time, rather than being
unalterable. Whilst it is important to have the three elements
controlled to a degree in a CoP, perseverance in one element
will help ease the potential problems in another. As Wenger et
al.\ have asserted, ``{\em The art of community development is
to use the synergy between domain, community, and practice to
help a community evolve and fulfil its potential.}'' \cite[page
47]{Wen02} Without the three elements above, a true CoP cannot
evolve. With them, the community can develop a {\em Body of
Knowledge} (BoK) that can be used by practitioners within a
particular area of expertise \cite{Bow11}.

A critical part of knowledge development is learning.
Increasingly it has been realised that this is largely
social in character, although it often takes place in the
workplace \cite{Har09}. In this framework, the concept of
legitimate peripheral participation (LPP) has been developed
\cite{Lav91}. This approach considers how individuals move from
being newcomers in a community, eventually becoming experienced
in some collaborative project or endeavour. Often the initial
tasks undertaken by participants are small-scale and low-risk.
Nevertheless, the act of empowering these peripheral members to
participate in a large-scale collaborative project promotes
interaction between novices and experts. It has the potential to
generate productive knowledge development within the community
involved in the overall effort. In the context of cybersecurity,
some people may have an initial interest because it has an
impact on their work. A proportion of these will go on to become
experts in the field, playing a leading role in the defence of
an organization that is critically dependent on networked IT for
example.

\section{Formalization in Z}
\label{formalization}

Communities have been studied in a variety of informal
frameworks, such as a Community of Practice as presented in the
previous section.  Typically there is some form of layered
structure to communities with sub-communities combining to form
larger communities. Here we suggest an abstract framework that
could be used to formulate the structure of communities of
people and associated sub-communities. A number of desirable
properties can be modelled. The framework is specified using the
Z notation \cite{Bow01,Spi01}, based on predicate logic and set
theory \cite{Hen03}, together with schema boxes for structuring
the mathematics forming the specification.  The choice of Z here
simply reflects the experience and background of the author,
although Z is particularly good at modelling relations, which
are helpful in this context for representing connections between
people and their associated communities.

In modelling communities, we initially define a given set,
$NAMES$ of entities, whether they are people or communities of
people.

\begin{zed}
[ NAMES ]
\end{zed}

The name space is split disjointly between people and
communities that provide structure for related people.

\begin{axdef}
PEOPLE, \\
COMMUNITIES : \power NAMES
\where
PEOPLE \cap COMMUNITIES = \emptyset
\end{axdef}

A basic community framework may be formulated as finite sets of
people and communities.

\def\COMzero{Community_0}

\begin{schema}{\COMzero}
people : \finset PEOPLE \\
communities : \finset COMMUNITIES
\end{schema}

Communities contain links between people.  The exact nature of
the links can be left open at this stage and can be fixed for
different situations. In the two examples presented in Section
\ref{visualization}, the links presented Facebook ``friendship''
and academic co-authorship respectively.  They could equally
well represent email contacts or other forms of communication,
for example.  Both people and communities may be members or a
part of other communities.  Links between people and community
membership should be valid. That is, links should relate actual
people in the community framework and all communities should
exist in the framework.

\def\COMone{Community_1}

\begin{schema}{\COMone}
\COMzero \\
links : PEOPLE \rel PEOPLE \\
memberships : NAMES \rel COMMUNITIES
\where
\dom links \subseteq people
\also
\dom memberships \subseteq people \cup communities
\also
\ran links \subseteq people
\also
\ran memberships \subseteq communities
\end{schema}

It is possible to specify people that have no links and entities
(people or communities) that are not within any community.
People may be ``orphans'' (i.e., have no links to them):

\def\COMtwo{Community_2}

\begin{schema}{\COMtwo}
\COMone \\
nolinks : \finset PEOPLE \\
nomemberships : \finset NAMES \\
orphans : \finset PEOPLE
\where
nolinks = people \setminus \dom links
\also
nomemberships = people \cup communities \setminus \dom memberships
\also
orphans = people \setminus \ran links
\end{schema}

\noindent
It may be desirable for all people to have links and be part of
(at least one) community and for all communities to have people
and/or sub-communities in them (i.e., for $nolinks$ and
$nomemberships$ to be empty). It may also be a desirable
property for there to be no orphans (i.e., for $orphans$ to be
empty).

We could strengthen earlier constraints. For example, we could
specify that all people have links with some other people and
are a member of some community. Also, all people are linked
from other people in some way and all communities are populated
(with memberships of people or sub-communities).

\def\COMthree{Community_3}

\begin{schema}{\COMthree}
\COMtwo
\where
people = \dom links
\also
people \subseteq \dom memberships
\also
\ran links = people
\also
\ran memberships = communities
\end{schema}

It is normally sensible to limit the model so that people cannot
be related to themselves and communities cannot be members of
themselves since this is not helpful for structuring.  Indeed,
loops are not desirable in categorising sub-communities, so it
is best to avoid transitive membership (indicated below by
$\plus$, irreflexive transitive closure).

\def\COMfour{Community_4}

\begin{schema}{\COMfour}
\COMtwo
\where
\id PEOPLE \cap links = \emptyset
\also
\id COMMUNITIES \cap memberships\plus = \emptyset
\end{schema}
 
There are some (one or more) top-level communities that are not
sub-communities of any other community. These top-level
communities provide one or more high-level starting points for
traversing the information about communities.

\def\COMfive{Community_5}

\begin{schema}{\COMfive}
\COMfour \\
toplevelcommunities : \finset_1 COMMUNITIES
\where
toplevelcommunities = \ran memberships \setminus \dom memberships
\end{schema}

If people are not interlinked in any way, it is questionable why
they are relevant in the overall community framework.

\def\COM{Community}

\begin{schema}{\COM}
\COMfive
\where
\dom links \cup \ran links = people
\end{schema}

The set of people associated with a particular person may be of
interest. We can define a status operation, where the state of
the overall community does not change.

\begin{schema}{LinkedPeople}
\Xi \COM \\
p? : PEOPLE \\
linked! : \finset PEOPLE
\where
linked! = links \limg \{ p? \} \rimg
\end{schema}

\noindent
(Note that the $\limg \ldots \rimg$ notation indicates the
relational image of a set.)

The set of common people associated with two specific people may
also be of interest:

\begin{schema}{CommonPeople}
\Xi \COM \\
p1?, p2? : PEOPLE \\
common! : \finset PEOPLE
\where
common! = links \limg \{ p1? \} \rimg \cap links \limg \{ p2? \} \rimg
\end{schema}

Information on the community membership of a person may be
wanted:

\begin{schema}{CommunityMembership}
\Xi \COM \\
p? : PEOPLE \\
communities! : \finset COMMUNITIES
\where
communities! = memberships\plus \limg \{ p? \} \rimg
\end{schema}

\noindent
Here all the different layers of community with which an
individual person is involved are returned.

The above Z specification has gradually built up a number of
desirable properties in a framework that could be used to
specify memberships of a layered set of communities together
with some example status operations on the overall community.
Further properties of communities could be added within this
model.  It is suggested that an abstract framework such as this
could be useful for formulating a conceptual model of
communities and used as a starting point for further reasoning
about and modelling of communities. This framework could be used
as a basis for visualization tools of communities.

\section{Conclusion}
\label{conclusion}

We have briefly considered visualization, development, and
formalization of communities of people using existing online
software tools, a Community of Practice framework, and the Z
notation respectively.  Visualising virtual communities has
become increasingly easy over the past decade as social and
professional networking has developed rapidly. Online software
tools are improving in this regard.  Communities of practice, as
postulated in the social science field, could be studied further
for a variety of communities \cite{Bow11,Liu11}. In particular,
visualization of these communities dynamically over time could
help in understanding their nature as they grow and contract.  Z
has proved to be an elegant formalism for capturing precise
descriptions of various aspects of connections between people
and associated entities due to its ability to model relations in
a natural way \cite{Bow13,Bow11}.

In summary, combining ideas from visualization and the Community
of Practice framework, underpinned by formalization in a
notation such as Z, could be any interesting area for further
exploration.  It can be expected that visualization of online
communities will improve significantly over the next decade just
as the communities themselves have developed and expanded
rapidly over the past decade. Further formalization could help
with understanding the nature of communities and the
relationships of people within them. Such studies could be
helpful in cybersecurity, with respect to online communities
associated with breaking security for example.

\paragraph*{Acknowledgements:}
Jonathan Bowen is grateful for financial support from Museophile
Limited.  The Z notation has been type-checked using the \fuzz\
type-checker \cite{Spi08}.

\newpage

\begin{footnotesize}

\end{footnotesize}

\end{document}